\documentclass[letterpaper,12pt]{JHEP3}
\usepackage{amsmath,amssymb}
\usepackage{latexsym}
\usepackage{epsfig}
\usepackage{cite}
\usepackage[english]{babel}

\newcommand{\eq}{\begin{equation}}
\newcommand{\feq}{\end{equation}}
\newcommand{\eqn}{\begin{eqnarray}}
\newcommand{\feqn}{\end{eqnarray}}
\newcommand{\arr}{\begin{eqnarray*}}
\newcommand{\farr}{\end{eqnarray*}}

\font\mybb=msbm10 at 12pt
\def\bb#1{\hbox{\mybb#1}}

\def\bR {\bb{R}}

\title{Conformal structure of the Schwarzschild black hole}

\author{Stefano Bertini,$^{ac}$ Sergio L.~Cacciatori$^{bc}$ and
Dietmar Klemm$^{ac}$ \\
$^a$ Dipartimento di Fisica dell'Universit\`a di Milano, \\
\hspace*{0.15cm} Via Celoria 16, I-20133 Milano. \\
$^b$ Dipartimento di Scienze Fisiche e Matematiche, \\
\hspace*{0.15cm} Universit\`a dell'Insubria, \\
\hspace*{0.15cm} Via Valleggio 11, I-22100 Como. \\
$^c$ INFN, Sezione di Milano, Via Celoria 16, I-20133 Milano. \\
}
\preprint{IFUM-978-FT}

\abstract{We show that the scalar wave equation at low frequencies in the
Schwarzschild geometry enjoys a hidden SL$(2,\bR)$ invariance, which is
not inherited from an underlying symmetry of the spacetime itself. Contrary to
what happens for Kerr black holes, the vector fields generating the SL$(2,\bR)$ are
globally defined. Furthermore, it turns out that under an SU$(2,1)$ Kinnersley transformation,
which maps the Schwarzschild solution into the near horizon limit $\text{AdS}_2\times\text{S}^2$
of the extremal Reissner-Nordstr\"om black hole (with the same entropy), the Schwarzschild
hidden symmetry generators become exactly the isometries of the AdS$_2$ factor.
Finally, we use the SL$(2,\bR)$ symmetry to determine algebraically the quasinormal frequencies
of the Schwarzschild black hole, and show that this yields the correct leading behaviour for
large damping.
}


\begin{document}

\section{Introduction}

The Schwarzschild solution in four spacetime dimensions represents perhaps the simplest
black hole at all, but it is nevertheless very difficult to understand its microstates. This is in
sharp contrast to some black holes in string theory, that (though being rather complicated)
can be understood in terms of bound states of D-branes and strings, which makes it possible
to compute their entropy microscopically \cite{Strominger:1996sh}.

A general feature that has emerged is that essentially all black holes whose entropy was
reproduced by a microstate counting, are described by two-dimensional conformal field theories.
Two of the most prominent examples of this type are the BTZ solution \cite{Strominger:1997eq}
or the extremal Kerr black hole \cite{Guica:2008mu}.

This universal conformal structure inherent to the physics of many black holes has triggered
numerous attempts to unveil such a conformal (albeit not necessarily two-dimensional) symmetry
also in the Schwarzschild case. For instance, \cite{Carlip:1998wz,Solodukhin:1998tc} considered diffeomorphisms preserving the black hole horizon, and showed that the charges associated to
these diffeomorphisms generate a centrally extended Virasoro algebra. A different approach was
adopted in \cite{Sachs:2001qb}, where it was shown that the optical metric of the Schwarzschild
solution becomes $\bR\times\text{H}^3$ near the horizon. Since $\text{H}^3$ is Euclidean
$\text{AdS}_3$, this allows to apply techniques originating from the AdS/CFT correspondence.
An interesting related development can be found in \cite{Barnich:2010eb}, where it was argued that
the symmetry algebra of asymptotically flat spacetimes at null infinity in four dimensions should be
taken to be the semi-direct sum of supertranslations with infinitesimal local conformal
transformations and not, as usually done, with the Lorentz algebra.

On the other hand, it was discovered recently \cite{Castro:2010fd} that the scalar wave equation in
the nonextremal Kerr black hole enjoys, in the low frequency limit,  a hidden conformal symmetry
that is not derived from an underlying symmetry of the spacetime itself. The existence of such a
hidden symmetry is related to the fact that black hole scattering amplitudes are given in terms of
hypergeometric functions \cite{Cvetic:1997xv}, which are well-known to form representations
of the conformal group $\text{SL}(2,\bR)$. Together with evidence provided by the results of
\cite{Guica:2008mu}, this led to the conjecture that the nonextremal Kerr
black hole with angular momentum $J$ is dual to a two-dimensional CFT with central charges
$c_L=c_R=12J$ \cite{Castro:2010fd}. Indeed, using $c_L$ and $c_R$ in the Cardy formula for
a CFT$_2$ gives exactly the Bekenstein-Hawking entropy of the Kerr solution. Moreover, the low
frequency scalar-Kerr scattering amplitudes coincide with thermal correlators of a 2d
CFT \cite{Castro:2010fd}\footnote{For the near-extremal Kerr black hole, this was first noted in
\cite{Cvetic:2009jn}.}.

In view of these results, one may ask whether an analogous hidden symmetry exists in the
Schwarzschild case as well. Note in this context that one cannot simply take the zero rotation
limit $(a\to 0)$ in the generators of the hidden $\text{SL}(2,\bR)\times\text{SL}(2,\bR)$ symmetry
for Kerr, since this limit is singular. For instance, the left- and right-temperatures of the dual CFT,
\eq
T_{\text L} = \frac{M^2}{2\pi J}\ , \qquad T_{\text R} = \frac{\sqrt{M^4 - J^2}}{2\pi J}\ ,
\feq
that appear in these generators, clearly diverge for $J\to 0$. We shall find that the massless
Klein-Gordon equation in the Schwarzschild background does indeed enjoy such a symmetry,
but with two essential differences to the Kerr case: First, there is only one $\text{SL}(2,\bR)$
factor present for the Schwarzschild black hole. Moreover, the generators are globally
defined, whereas for Kerr they are not periodic under the angular identification $\phi\sim\phi+2\pi$,
which breaks $\text{SL}(2,\bR)_{\text L}\times\text{SL}(2,\bR)_{\text R}$ down to
$\text{U}(1)_{\text L}\times\text{U}(1)_{\text R}$.

The remainder of this paper is organized as follows: In section \ref{sect-hidden}, we describe
the hidden $\text{SL}(2,\bR)$ symmetry appearing in the scalar wave equation at low
frequencies. In the following section, we use the fact that the four-dimensional stationary
Einstein-Maxwell equations are invariant under an $\text{SU}(2,1)$ group of
transformations \cite{Neugebauer:1969wr,Kinnersley:1973} to map the Schwarzschild solution
into the near-horizon limit $\text{AdS}_2\times\text{S}^2$ of the extremal Reissner-Nordstr\"om black
hole. We show that this transformation preserves the entropy, and makes the hidden $\text{SL}(2,\bR)$
symmetry manifest, since its generators become exactly the isometries of the AdS$_2$ factor.
After that, in section \ref{quasi}, the $\text{SL}(2,\bR)$ symmetry is used to algebraically determine
the Schwarzschild quasinormal modes as descendents of a lowest weight state. Although this
takes us out of the validity of our low (and real) frequency approximation, it surprisingly yields
the correct leading behaviour of the quasinormal frequencies for large damping.
We conclude in \ref{final} with some final remarks. In appendix \ref{app-higherd} it is shown
that the hidden $\text{SL}(2,\bR)$ symmetry extends to the Schwarzschild black hole in any
dimension.

\section{Hidden conformal symmetry}
\label{sect-hidden}

Let us consider the massless Klein-Gordon equation
\eq
\frac1{\sqrt{-g}}\partial_{\mu}(\sqrt{-g}g^{\mu\nu}\partial_{\nu}\Phi) = 0 \label{KG}
\feq
in the Schwarzschild geometry,
\eq
ds^2 = -V(r)dt^2 + \frac{dr^2}{V(r)} + r^2(d\theta^2 + \sin^2\theta d\phi^2)\ , \qquad
V(r) = 1 - \frac{2M}r\ .
\feq
Using the separation ansatz
\eq
\Phi(t,r,\theta,\phi) = e^{-i\omega t}R(r)Y^l_m(\theta,\phi)\ ,
\feq
together with
\begin{displaymath}
\Delta_{S^2}Y^l_m(\theta,\phi) = \frac1{\sin\theta}\partial_{\theta}(\sin\theta\partial_{\theta}
Y^l_m(\theta,\phi)) + \frac1{\sin^2\theta}\partial^2_{\phi}Y^l_m(\theta,\phi) = -l(l+1)Y^l_m(\theta,\phi)\ ,
\end{displaymath}
\eqref{KG} reduces to
\eq
\partial_r\Delta\partial_rR + \frac{\omega^2r^4}{\Delta}R - l(l+1)R = 0\ , \label{radial}
\feq
where we defined $\Delta=r^2-2Mr\equiv r(r-r_+)$. Now use
\begin{displaymath}
\frac{\omega^2r^4}{\Delta} = \omega^2r^2 + \omega^2rr_+ + \omega^2r_+^2 +
\frac{\omega^2r_+^3}r + \frac{\omega^2r_+^4}{\Delta}\ .
\end{displaymath}
The first four terms on the rhs are much smaller than $1$ in the near-region, low frequency
limit $\omega r\ll 1$, $\omega r_+\ll 1$, whereas the last term blows up if one
goes sufficiently close to the horizon. We shall thus approximate\footnote{Note that this
approximation is exactly the same as the one in eqns.~(2.11), (2.12) of
\cite{Maldacena:1997ih} (set $a=Q=0$ there). The authors of \cite{Maldacena:1997ih}
have $r-r_+\ll 1/\omega$ as near-region condition, which is implied by
$\omega r\ll 1$, $\omega r_+\ll 1$ and $r>r_+$.} the expression
$\omega^2r^4/\Delta$ by $\omega^2r_+^4/\Delta$. Then \eqref{radial} becomes
\eq
\partial_r\Delta\partial_rR + \frac{\omega^2r_+^4}{\Delta}R - l(l+1)R = 0\ . \label{radial-approx}
\feq
Next, we define the vector fields
\begin{eqnarray}
H_1 &=& ie^{\frac t{4M}}\left(\Delta^{1/2}\partial_r - 4M(r-M)\Delta^{-1/2}\partial_t\right)\ , \nonumber \\
H_0 &=& -4iM\partial_t\ , \label{vec-fields} \\
H_{-1} &=& -ie^{-\frac t{4M}}\left(\Delta^{1/2}\partial_r + 4M(r-M)\Delta^{-1/2}\partial_t\right)\ , \nonumber
\end{eqnarray}
which obey the SL$(2,\bR)$ commutation relations
\eq
[H_0,H_{\pm 1}] = \mp iH_{\pm 1}\ , \qquad [H_1,H_{-1}] = 2iH_0\ . \label{sl2R}
\feq
The SL$(2,\bR)$ Casimir reads
\begin{eqnarray}
{\cal H}^2 &=& -H_0^2 + \frac12(H_1H_{-1} + H_{-1}H_1) \nonumber \\
&=& \Delta\partial^2_r + 2(r-M)\partial_r - \frac{16M^4}{\Delta}\partial^2_t\ ,
\end{eqnarray}
and thus the near-region Klein-Gordon equation can be rewritten as
\eq
{\cal H}^2\Phi = l(l+1)\Phi\ . \label{KG-algebraic}
\feq
We see that the
scalar wave equation in the Schwarzschild geometry enjoys a hidden conformal
symmetry similar to the Kerr case, but with two essential differences: First, for the
Kerr black hole, there is an $\text{SL}(2,\bR)_{\text L}\times\text{SL}(2,\bR)_{\text R}$
symmetry, whereas here only one SL$(2,\bR)$ factor is present. This indicates that
the Schwarzschild black hole might be described by a chiral CFT.
Secondly, the vector fields \eqref{vec-fields} are globally defined, while
the ones in \cite{Castro:2010fd} are not periodic under the angular identification $\phi\sim\phi+2\pi$.
This fact was interpreted in \cite{Castro:2010fd} as a spontaneous breaking of the
$\text{SL}(2,\bR)_{\text L}\times\text{SL}(2,\bR)_{\text R}$ symmetry down to
$\text{U}(1)_{\text L}\times\text{U}(1)_{\text R}$ by left and right temperatures $T_{\text L}$
and $T_{\text R}$.

Notice also that the existence of a hidden $\text{SL}(2,\bR)$ symmetry stems from
the fact that the solution of \eqref{radial} is given in terms of Heun functions,
which can be expanded in a series of hypergeometric functions (cf.~e.g.~\cite{Heun}),
and the latter form representations of the conformal group. (One can show that this series
can be truncated to the leading term in a low energy limit, cf.~\cite{Lowe:2011wu} for
a review of such an expansion for the Kerr black hole).

The condition \eqref{KG-algebraic} implies that the field $\Phi$ has conformal weight
$h=l+1$ \cite{Lowe:2011wu}. To see this, define $L_n=-iH_n$, $n=0,\pm 1$, such that
\eq
[L_n, L_m] = (n-m)L_{n+m}\ .
\feq
Then the second Casimir is
\eq
{\cal H}^2 = L_0^2 - \frac12\left(L_1L_{-1} + L_{-1}L_1\right) = h(h-1)\ ,
\feq
which implies $h(h-1)=l(l+1)$, whose positive solution is $h=l+1$.

Note that a similar conformal structure was discovered before
in \cite{Govindarajan:2000ag} and further explored in \cite{Birmingham:2001qa,Gupta:2001bg}:
If we set $R=\chi/\Delta^{1/2}$, the radial equation \eqref{radial} reduces to
\eq
\partial^2_r\chi + \frac{M^2 + \omega^2r^4}{\Delta^2}\chi - \frac{l(l+1)}{\Delta}\chi = 0\ .
\feq
Defining $x=r-r_+$ and expanding the potential near the horizon $x=0$, this boils
down to
\eq
\left(-\frac{d^2}{dx^2} - \frac g{4x^2} + {\cal O}(x^{-1})\right)\chi = 0\ , \label{DFF}
\feq
where $g=1+(4M\omega)^2$. The operator
\eq
H = -\frac{d^2}{dx^2} - \frac g{4x^2}
\feq
is nothing else than the Hamiltonian of the De Alfaro-Fubini-Furlan (DFF) model of conformal
quantum mechanics \cite{deAlfaro:1976je}. $H$, together with
\eq
D = \frac i4\left(x\frac d{dx}+\frac d{dx}x\right)\ , \qquad K = \frac 14x^2\ ,
\feq
generating dilatations and special conformal transformations respectively,
satisfy the sl$(2,\bR)$ algebra
\eq
[D,H] = -iH\ , \qquad [D,K] = iK\ , \qquad [H,K] = 2iD\ .
\feq
While it is well-known that the dynamics of a particle near the horizon of an extremal
Reissner-Nordstr\"om black hole is governed by a model of conformal
mechanics \cite{Claus:1998ts} (this is just a consequence of the SL$(2,\bR)$
isometry group of the AdS$_2$ factor contained in the near-horizon geometry), the
appearance of the DFF model for the Schwarzschild black hole is less obvious.
In this case, the conformal symmetry is hidden, i.e., it is not inherited from a near-horizon
geometry that has this symmetry.

\section{Kinnersley transformations and relation to $\text{AdS}_2\times\text{S}^2$}
\label{kinnersley}

It is well-known that the four-dimensional stationary Einstein-Maxwell equations are invariant
under an SU$(2,1)$ group of transformations \cite{Neugebauer:1969wr,Kinnersley:1973}.
In this section, we shall use this group to map the Schwarzschild solution  to the
$\text{AdS}_2\times\text{S}^2$ Bertotti-Robinson spacetime, and show that the hidden
symmetry generators \eqref{vec-fields} go over into the AdS$_2$ isometries under this mapping.

We first review briefly how the SU$(2,1)$ acts on the solution space of the stationary
Einstein-Maxwell equations. Any spacetime admitting a timelike Killing vector can be
written as\footnote{In order to conform to some of the older literature on this subject, in this
section (and only here) we use mostly minus signature.}
\eq
ds^2 = f(dt - \omega_i dx^i)^2 - f^{-1}h_{ij}dx^i dx^j\ , \label{stationary}
\feq
where the scalar $f$, the one-form $\omega_i$ and the three-metric $h_{ij}$ depend
on the spatial coordinates $x^i$ only. The electromagnetic field $F_{\mu\nu}$ can be
parametrized in terms of electric and magnetic potentials $u$ and $v$,
\eq
F_{i0} = \partial_i v\ , \qquad F^{ij} = fh^{-1/2}\epsilon^{ijk}\partial_k u\ .
\feq
Moreover, one defines the twist or nut-potential $\chi$ by
\eq
\partial_i\chi = -f^2h^{-1/2}h_{ij}\epsilon^{jkl}\partial_k\omega_l + 2(u\partial_i v - v\partial_i u)\ ,
\feq
and combines the four real scalars $f,\chi,u,v$ to the complex Ernst potentials according to
\eq
{\cal E}Ê= f + i\chi - \bar\psi\psi\ , \qquad \psi = v + iu\ .
\feq
Then the stationary Einstein-Maxwell equations boil down to \cite{Harrison:1968,Neugebauer:1969wr}
\begin{eqnarray}
f\nabla^2{\cal E} &=& \nabla{\cal E}\cdot\left(\nabla{\cal E} + 2\bar\psi\nabla\psi\right)\ , \nonumber \\
f\nabla^2\psi &=& \nabla\psi\cdot\left(\nabla{\cal E} + 2\bar\psi\nabla\psi\right)\ , \label{Ernst} \\
f^2 R_{ij}(h) &=& \text{Re}\left[\frac12{\cal E}_{,\left(i\right.}\bar{\cal E}_{,\left.j\right)} +
2\psi{\cal E}_{,\left(i\right.}\bar\psi_{,\left.j\right)} - 2{\cal E}\psi_{,\left(i\right.}\bar\psi_{,\left.j\right)}
\right]\ , \nonumber
\end{eqnarray}
where the scalar products and the Laplacian are computed with the metric $h_{ij}$.
The equations \eqref{Ernst} are invariant under an SU$(2,1)$ group of
transformations \cite{Neugebauer:1969wr,Kinnersley:1973}, acting as follows: Parametrize the
Ernst potentials in terms of the three Kinnersley potentials $U,V,W$ (one of which is redundant)
by \cite{Kinnersley:1973}
\eq
{\cal E} = \frac{U - W}{U + W}\ , \qquad \psi = \frac V{U + W}\ .
\feq
Then, the SU$(2,1)$ acts linearly on the complex vector $(U,V,W)$, and transforms solutions
of \eqref{Ernst} with spatial metric $h_{ij}$ into new solutions with the same $h_{ij}$.
Note that the SU$(2,1)$ invariance is just a consequence of the fact that the timelike Kaluza-Klein
reduction of the 4d Einstein-Maxwell action yields three-dimensional gravity coupled to an
$\text{SU}(2,1)/\text{S}(\text{U}(1,1)\times\text{U}(1))$ nonlinear sigma model, which describes the
four scalars $f,\chi,u,v$ \cite{Breitenlohner:1987dg}\footnote{This extends also to generalizations
of the Einstein-Maxwell action which typically arise from Kaluza-Klein theories. In that case one
gets more complicated $G/H$ nonlinear sigma models \cite{Breitenlohner:1987dg}.}.

In order to apply this to the Schwarzschild solution, rewrite the latter as
\eq
ds^2 = f dt^2 - f^{-1}M^2\left[dx^2 + (x^2 - 1)(d\theta^2 + \sin^2\theta d\phi^2)\right]\ , \qquad
f = \frac{x-1}{x+1}\ ,
\feq
where the new coordinate $x$ is given by $x=r/M-1$, such that the horizon is at $x=1$.
As Ernst potentials we may thus take $U=x$, $V=0$, $W=1$. We now apply a boost in the
$(V,W)$ subspace, followed by an involution in $U,V$,
\eq
\left(\begin{array}{c} V' \\ W' \end{array}\right) = \left(\begin{array}{cc} \cosh\alpha & \sinh\alpha \\
\sinh\alpha & \cosh\alpha \end{array}\right)\left(\begin{array}{c} V \\ W \end{array}\right)\ , \qquad
U' = U\ , \label{boost}
\feq
\eq
U'' = V'\ , \qquad V'' = U'\ , \qquad W'' = W'\ ,
\feq
which leads to the new metric\footnote{Notice that the boost \eqref{boost} alone maps the
Schwarzschild into the nonextremal Reissner-Nordstr\"om solution, cf.~e.g.~\cite{Clement:1997tx}.}
\eq
{ds''}^2 = e^{-2\alpha}(x^2 - 1)dt^2 - \frac{e^{2\alpha}M^2dx^2}{x^2 - 1} - e^{2\alpha}M^2
(d\theta^2 + \sin^2\theta d\phi^2)\ , \label{new-metr}
\feq
and gauge field
\eq
F'' = e^{-\alpha}dx\wedge dt\ .
\feq
\eqref{new-metr} is the Bertotti-Robinson spacetime $\text{AdS}_2\times\text{S}^2$,
with the AdS$_2$ seen by an accelerated observer, and $x=1$ the acceleration horizon.
It represents the near-horizon geometry of the extremal Reissner-Nordstr\"om black hole,
with entropy
\eq
S = \frac{A_{\text{hor}}}{4G} = \frac{e^{2\alpha}\pi M^2}{G}\ .
\feq
Apparently, this is different from the entropy $4\pi M^2/G$ of the Schwarzschild black hole
we started with, unless $e^{\alpha}=2$. However, there is a subtlety here: Consider the
timelike Kaluza-Klein reduction from four to three dimensions, using the ansatz \eqref{stationary},
and Wick-rotate $t=-i\tau$, with $\tau\sim\tau + \beta$, where $\beta$ denotes the inverse
temperature. This yields an effective three-dimensional Newton constant $G_3=G_4/\beta$.
If the Kinnersley transformation maps a solution with Euclidean time period $\beta$ into
one with $\hat\beta$, we have obviously
\eq
\frac1{G_3} = \frac{\beta}{G_4} = \frac{\hat\beta}{\hat G_4}\ .
\feq
The entropy of the new solution is thus
\eq
\hat S = \frac{\hat A_{\text{hor}}}{4\hat G_4} = \frac{\hat A_{\text{hor}}\beta}{4G_4\hat\beta}\ . \label{hatS}
\feq
In our case, the inverse temperature associated to the horizon at $x=1$ of \eqref{new-metr} is
easily seen to be $\hat\beta=2\pi Me^{2\alpha}$, whereas $\beta=8\pi M$ for Schwarzschild.
Since $\hat A_{\text{hor}}=e^{2\alpha}A_{\text{hor}}/4$, \eqref{hatS} gives $\hat S=S$, so that
the entropy is actually invariant, no matter what the value of the boost parameter $\alpha$ is.
We are not aware of any general proof that Kinnersley transformations leave the Bekenstein-Hawking
entropy invariant, as it happens e.g.~for T-duality in string theory \cite{Horowitz:1993wt} (it is not even
evident that they map black holes into solutions that have again a horizon), but in our special case they obviously do. Notice that the value $e^{\alpha}=2$ is nevertheless special, in that the temperature
of \eqref{new-metr} coincides exactly with the temperature of the Schwarzschild black hole.
Moreover, the SL$(2,\bR)$ generators \eqref{vec-fields}, which in the coordinate $x$ read
\begin{eqnarray}
H_1 &=& ie^{\frac t{4M}}\left((x^2-1)^{1/2}\partial_x - 4Mx(x^2-1)^{-1/2}\partial_t\right)\ , \nonumber \\
H_0 &=& -4iM\partial_t\ , \\
H_{-1} &=& -ie^{-\frac t{4M}}\left((x^2-1)^{1/2}\partial_x + 4Mx(x^2-1)^{-1/2}\partial_t\right)\ , \nonumber
\end{eqnarray}
are exactly the Killing vectors of the AdS$_2$ factor in \eqref{new-metr} for $e^{\alpha}=2$. (For
other values of the boost parameter, one has to rescale time in order to have this identification).
Also, the near-region, low frequency Klein-Gordon equation \eqref{KG-algebraic} becomes
precisely the KG equation on $\text{AdS}_2\times\text{S}^2$.

\section{Quasinormal modes}
\label{quasi}

Quasinormal modes \cite{Nollert:1999ji} are defined to be perturbations of the black hole whose
boundary conditions are purely outgoing both at the horizon and at infinity. These boundary
conditions single out discrete complex frequencies $\omega_n$. It has been argued
\cite{Dreyer:2002vy} that the asymptotic (large $n$) behaviour of the high overtone black hole
quasinormal frequencies captures important information about the spectrum of black hole observables;
in particular that the asymptotic value of $\text{Re}\,\omega_n$ is related to the so-called Barbero-Immirzi parameter of loop quantum gravity.

In general, the $\omega_n$ have to be determined numerically, for instance by using continued
fraction techniques. In this way, Nollert \cite{Nollert:1993zz} obtained
\eq
M\omega_n = 0.0437123 - \frac i4\left(n + \frac12\right) + {\cal O}\left[(n + 1)^{-1/2}\right]
\label{quasi-Schwarz}
\feq
for the scalar quasinormal modes of the Schwarzschild black hole. It was first realized by
Hod \cite{Hod:1998vk} that the numerical value $0.0437123$ agrees (up to the available precision) with
$\ln 3/(8\pi)$, a number required by statistical physics arguments and Bohr's correspondence
principle. Later it was shown in \cite{Motl:2002hd,Motl:2003cd} that the asymptotic real part of
$\omega_n$ is indeed precisely $\ln 3/(8\pi)$.

The authors of \cite{Chen:2010ik} realized that in black hole spacetimes with hidden conformal
symmetry, one can use the latter to algebraically determine the quasinormal mode spectrum
as descendents of a lowest weight state. Looking at \eqref{quasi-Schwarz}, we see that
for large $n$ the imaginary part of $\omega_n$ is equally spaced\footnote{Note that the
spacing $2\pi iT_{\text{Hawking}}$ in \eqref{quasi-Schwarz} is not too surprising, since the
quasinormal modes determine the position of poles of a Green's function, and the black hole
has Euclidean time $\tau\sim\tau+1/T_{\text{Hawking}}$ \cite{Motl:2003cd}.}, so one might ask
whether this can be realized as an SL$(2,\bR)$ tower. Apparently the limit of large
$\text{Im}\,\omega_n$ takes us out of the validity of our approximation. (In order to reduce the
Klein-Gordon operator to an SL$(2,\bR)$ Casimir we used $\omega$ real and $\omega M\ll1$).
One might nevertheless ask how far the applicability of the SL$(2,\bR)$ symmetry can be pushed,
and see which quasinormal modes result by acting with $L_{-1}$ on a lowest weight state.
We shall see that this reproduces correctly the leading large $n$ behaviour of \eqref{quasi-Schwarz}.

Let us denote the lowest weight state by $\Phi^{(0)}$. By definition
\eq
L_0\Phi^{(0)} = h\Phi^{(0)}\ , \qquad L_1\Phi^{(0)} = 0\ , \label{highest}
\feq
where $L_m=-iH_m$. Since
\eq
\Phi^{(0)} = e^{-i\omega_0 t}R^{(0)}(r)Y^l_m(\theta,\phi)\ , \label{Phi0}
\feq
we have $h=4iM\omega_0$. Using \eqref{highest} together with $L_1L_{-1}=2L_0+L_{-1}L_1$
in \eqref{KG-algebraic}, one gets
\eq
h = \frac 12\left(1 \pm (2l + 1)\right)\ ,
\feq
and thus
\eq
M\omega_0 = -\frac i8\left(1 \pm (2l + 1)\right)\ .
\feq
Since quasinormal modes have $\text{Im}\,\omega_n<0$ we must choose the upper sign,
such that $M\omega_0=-i(l+1)/4$ and $h=l+1$, in agreement with the conformal weight
assignment in section \ref{sect-hidden}. One can now construct the descendents
\eq
\Phi^{(n)} = (L_{-1})^n\Phi^{(0)}\ .
\feq
Taking into account \eqref{Phi0} as well as $L_{-1}=-iH_{-1}$, with the expression for $H_{-1}$
given in eqns.~\eqref{vec-fields}, it is not difficult to shew that
\eq
\Phi^{(n)} = e^{-i\omega_n t}R^{(n)}(r)Y^l_m(\theta,\phi)\ ,
\feq
where
\eq
M\omega_n = M\omega_0 - \frac i4n\ , \label{quasi-algebraic}
\feq
and
\begin{eqnarray}
R^{(n)}(r) = &&\left(-\Delta^{1/2}\partial_r + 4M(r-M)\Delta^{-1/2}i\omega_{n-1}\right)
\left(-\Delta^{1/2}\partial_r + 4M(r-M)\Delta^{-1/2}i\omega_{n-2}\right)\cdot \nonumber \\
&&\ldots\cdot\left(-\Delta^{1/2}\partial_r + 4M(r-M)\Delta^{-1/2}i\omega_0\right)R^{(0)}(r)\ .
\end{eqnarray}
Notice that \eqref{quasi-algebraic} implies
\eq
L_0\Phi^{(n)} = (h+n)\Phi^{(n)}\ , \qquad n = 0,1,\ldots\ ,
\feq
and thus the quasinormal modes $\Phi^{(n)}$ form a principal discrete lowest weight
representation of $\text{SL}(2,\bR)$.

Comparing \eqref{quasi-algebraic} with \eqref{quasi-Schwarz}, we see that the leading
behaviour $\sim -in/4$ for large damping comes out correctly, while the subleading terms
do not. In particular, the frequencies \eqref{quasi-algebraic} are purely imaginary.

Let us finally check if the $\Phi^{(n)}$ satisfy purely outgoing boundary conditions at the
horizon, as it must be for quasinormal modes. We have to show that
\eq
R^{(n)} \sim e^{-i\omega_n r_{\star}} \quad \text{as} \quad r_{\star}\to -\infty\ , \label{behav-R^n}
\feq
where $r_{\star}$ is the tortoise coordinate
\eq
r_{\star} = r + 2M\ln\left(\frac r{2M} - 1\right)\ . \label{tortoise}
\feq
First of all, $L_1\Phi^{(0)}=0$ yields
\eq
R^{(0)} = C(r^2 - 2Mr)^{-2iM\omega_0}\ ,
\feq
with $C$ an integration constant. This is easily seen to behave as
\eq
R^{(0)} \sim e^{-i\omega_0r_{\star}}\left(1 + {\cal O}(e^{r_{\star}/(2M)})\right)
\feq
as $r_{\star}\to-\infty$. Let us show \eqref{behav-R^n} by induction. To this end, assume that
\eq
R^{(n-1)} \sim e^{-i\omega_{n-1}r_{\star}}\left(1 + {\cal O}(e^{r_{\star}/(2M)})\right) \quad
\text{as} \quad r_{\star}\to -\infty\ , \label{Ind-Annahme}
\feq
which clearly holds for $n=1$. Acting on \eqref{Ind-Annahme} with the operator
$(-\Delta^{1/2}\partial_r + 4M(r-M)\Delta^{-1/2}i\omega_{n-1})$, one finds that
\eqref{Ind-Annahme} holds also with $n-1$ replaced by $n$, which proves \eqref{behav-R^n}.
Note that the functions $R^{(n)}(r)$ do not satisfy outgoing boundary conditions at infinity,
but this was to be expected, since they are solutions only in some near region, and have to
be matched somewhere with a far region solution.

In view of the results of this section, it would be very interesting to see if one can set up a
perturbation expansion organized in terms of the $\text{SL}(2,\bR)$ that gives the frequencies
\eqref{quasi-Schwarz}. Work in this direction is in progress.

\section{Final remarks}
\label{final}

Our results indicate that the Schwarzschild black hole might have a description in terms
of a two-dimensional CFT. If this is the case, the $\text{SL}(2,\bR)$ \eqref{sl2R} should be
enlarged to the whole Virasoro algebra, which raises the question if the corresponding
generators are related to those of \cite{Carlip:1998wz,Solodukhin:1998tc}, that generate
diffeomorphisms preserving certain boundary conditions at the black hole horizon.
In this context, it is interesting to consider
the transformation from Schwarzschild to Kruskal coordinates $U,V$, given by
\eq
U = -e^{-u/4M}\ , \qquad V = e^{v/4M}\ , \label{kruskal}
\feq
where $u=t-r_{\star}$, $v=t+r_{\star}$, and $r_{\star}$ denotes the tortoise coordinate
defined in \eqref{tortoise}.
In the Euclidean section we have $t=-i\tau$, where $\tau$ is identified modulo $\beta=1/T=8\pi M$.
Defining $w=r_{\star}+i\tau$ and $z=-U$, \eqref{kruskal} becomes
\eq
z = e^{\frac w{4M}}\ .
\feq
This is exactly the conformal transformation from a cylinder ($w$) to a plane ($z$), namely
$z=\exp(2\pi w/L)$, if the circumference $L$ of the cylinder is identified with the inverse
temperature $\beta$. It is well-known that such a transformation induces a shift of $c/24$ in the
Virasoro generators $L_0,\tilde L_0$. If we knew how to define the stress tensor of the dual
CFT (in a way similar to that of the AdS/CFT correspondence \cite{Balasubramanian:1999re}),
this would allow to compute the central charge $c$.

Notice also that the Schwarzschild solution is related by the duality-type transformation of section
\ref{kinnersley} to the near horizon limit $\text{AdS}_2\times\text{S}^2$ of the extremal
Reissner-Nordstr\"om black hole, and the latter is known to be described by a
CFT$_2$ \cite{Hartman:2008pb}. It would be interesting to see what the Kerr solution maps to under
this $\text{SU}(2,1)$ transformation, and if (part of) the hidden conformal
$\text{SL}(2,\bR)\times\text{SL}(2,\bR)$ symmetry of \cite{Castro:2010fd} becomes manifest in this way.

An open question (also in Kerr/CFT) is the massive case: In the AdS/CFT correspondence, a
mass term for a bulk field modifies the conformal weight of the dual operator. In order to see
whether something similar happens here (or in Kerr/CFT), one would have to show that the massive
Klein-Gordon equation still enjoys a hidden conformal symmetry, but now with shifted
weight for $\Phi$. For $m\neq 0$ there is an additional term $-m^2r^2R$ on the lhs of
\eqref{radial-approx}. Since
\begin{displaymath}
r^2 = r_+^2 + 2r_+(r - r_+) + (r - r_+)^2\ ,
\end{displaymath}
one can approximate $m^2r^2$ by $m^2r_+^2$ provided that $r-r_+\ll r_+$. Then, everything
goes through as before, with $l(l+1)$ replaced by $l(l+1)+m^2r_+^2=h(h-1)$, so that now
$\Phi$ has weight
\eq
h = \frac12\left[1 + \sqrt{(1+2l)^2 + (4Mm)^2}\right]\ .
\feq
Note that, contrary to the approximation that reduces the Klein-Gordon operator to an
$\text{SL}(2,\bR)$ Casimir used in the massless case, the replacement of $m^2r^2$ by
$m^2r_+^2$ is a true near-horizon limit. A more detailed study of the massive case,
as well as an investigation if the $\text{SL}(2,\bR)$ of section \ref{sect-hidden}
extends also to fields of nonvanishing spin, will be presented elsewhere.

\acknowledgments

We would like to thank Angela Mandelli for the caff\`e d'orzo and for
the nice atmosphere which has led to this paper. D.~K.~also thanks G.~Barnich and
A.~Virmani for discussion. This work was partially supported by INFN.

\normalsize

\appendix

\section{SL$(2,\bR)$ symmetry in $d$ dimensions}
\label{app-higherd}

In $d$ dimensions the Schwarzschild solution reads
\begin{eqnarray}
&& ds^2 = -V(r) dt^2 + \frac {dr^2}{V(r)} + r^2 d\Omega_{d-2}^2\ , \\
&& V(r)=1-\frac {r_+^{d-3}}{r^{d-3}}, \qquad\ r_+=\frac {8\pi \Gamma ((d-2)/2)}{(d-2)\pi^{\frac {d-1}2}} M\ .
\nonumber
\end{eqnarray}
Using the separation ansatz
\begin{equation}
\Phi(t,r,\vec \theta)= e^{-i\omega t} R(r) Y^l_\mu (\vec \theta)\ ,
\end{equation}
where $Y^l_\mu$ are the spherical harmonics on $S^{d-2}$, the massless Klein-Gordon equation
for $\Phi$ becomes
\begin{eqnarray}
\frac {\omega^2}{V} R+\frac 1{r^{d-2}} \partial_r \left( r^{d-2} V \partial_r R \right) -\frac {l(l+d-3)}{r^2} R=0\ .
\label{KG-d}
\end{eqnarray}
To recover the hidden symmetry, let us first introduce the change of variables
\begin{equation}
\rho=r^{d-3}\ ,
\end{equation}
so that \eqref{KG-d} takes the form
\begin{eqnarray}
&& \partial_\rho (\Delta_\rho \partial_\rho R) +\frac {\omega^2 r^2}{(d-3)^2 V} R -\frac {l(l+d-3)}{(d-3)^2} R=0\  ,
\label{radial-d} \\
&& \Delta_\rho=\rho(\rho-r_+^{d-3})\ . \nonumber
\end{eqnarray}
Using the identity
\begin{displaymath}
\frac {\omega^2 r^2}{(d-3)^2 V}
%=\frac {\omega^2 r^2}{(d-3)^2} \frac {\rho^2}{\rho(\rho-\rho_+)}=
%\frac {\omega^2 r^2}{(d-3)^2}+\frac {\omega^2 r^2}{(d-3)^2} \frac {\rho_+}{\rho}+\frac {\omega^2 r^2
%}{(d-3)^2} \frac {\rho_+^2}{\rho (\rho-\rho_+)}\cr
%&&=\frac {\omega^2 r^2}{(d-3)^2}+\frac {\omega^2 r^2}{(d-3)^2} \frac {\rho_+}{\rho}+\frac {\omega^2 
%r_+^2}{(d-3)^2} \frac {\rho_+^2}{\rho(\rho-\rho_+)}
%+\frac {\omega^2}{(d-3)^2} \frac {\rho_+^2}{\rho} \frac {r^2-r_+^2}{(r^{d-3}-r_+^{d-3})}\cr
%&&=\frac {\omega^2 r^2}{(d-3)^2}+\frac {\omega^2 r^2}{(d-3)^2} \frac {\rho_+}{\rho}+\frac {\omega^2 
%r_+^2}{(d-3)^2} \frac {\rho_+^2}{\rho(\rho-\rho_+)}
%+\frac {\omega^2}{(d-3)^2} \frac {\rho_+^2}{\rho} \frac {r+r_+}{\sum_{i=0}^{d-4} r^i r_+^{d-4-i}}\cr
=\frac {\omega^2 r^2}{(d-3)^2} \left[ 1+ \left(\frac {r_+}r\right)^{d-3} +\left(\frac {r_+}r\right)^{d-2} 
\left(1+\frac {r_+}r\right) \frac 1{\sum_{i=0}^{d-4} \left(\frac {r_+}r\right)^i } \right]+\frac {\omega^2 r_+^{2d-4}}
{(d-3)^2 \Delta_\rho}\ ,
\end{displaymath}
we see that the expression $r^2/V$ can be approximated by $r_+^{2d-4}/\Delta_{\rho}$
in the near-region, low frequency limit $\omega r\ll 1$, $\omega r_+\ll 1$. Then \eqref{radial-d} becomes
\begin{eqnarray}
&& \partial_\rho (\Delta_\rho \partial_\rho R) +\frac {\omega^2 r_+^{2d-4}}{(d-3)^2 \Delta_\rho} R -\frac {l(l+d-3)}{(d-3)^2} R=0\ ,
\end{eqnarray}
which has the same structure as (\ref{radial-approx}).
The vector fields
\begin{eqnarray}
&& H_1= ie^{\frac {(d-3)t}{2r_+}} \left(\Delta_\rho^{1/2} \partial_\rho -\frac {r_+}{d-3} (2\rho-r_+^{d-3})
\Delta_\rho^{-1/2}\partial_t \right)\ , \nonumber \\
&& H_0= -i\frac {2r_+}{d-3}\partial_t\ , \\
&& H_{-1}= -ie^{\frac {-(d-3)t}{2r_+}} \left(\Delta_\rho^{1/2} \partial_\rho +\frac {r_+}{d-3} (2\rho-r_+^{d-3})
\Delta_\rho^{-1/2}\partial_t \right)\ , \nonumber
\end{eqnarray}
satisfy the SL$(2,\bR)$ commutation relations
\eq
[H_0,H_{\pm 1}] = \mp iH_{\pm 1}\ , \qquad [H_1,H_{-1}] = 2iH_0\ ,
\feq
and the corresponding Casimir reads
\begin{eqnarray}
{\cal H}^2 &=& -H_0^2 + \frac12(H_1H_{-1} + H_{-1}H_1) \nonumber \\
&=& \Delta_\rho\partial^2_\rho + (2\rho-r_+^{d-3})\partial_\rho - \frac {r_+^{2d-4}}{(d-3)^2
\Delta_\rho}\partial^2_t\ ,
\end{eqnarray}
so that the near-region, low frequency Klein-Gordon equation takes the form
\begin{eqnarray}
{\cal H}^2 \Phi=\tilde l(\tilde l+1)\Phi\ , \qquad \tilde l=\frac l{d-3}\ ,
\end{eqnarray}
which implies that the field $\Phi$ has conformal weight $h=\tilde l + 1$.
Going back to the $r$ coordinate and defining $\Delta=r(r^{d-3}-r_+^{d-3})$, we get
\begin{eqnarray}
&& H_1= \frac i{(d-3)r^{\frac d2-2}} e^{\frac {(d-3)t}{2r_+}} \left(\Delta^{1/2} \partial_r
-\frac {r_+ (2r^{d-3}-r_+^{d-3})}{\Delta^{1/2}} \partial_t \right)\ , \nonumber \\
&& H_0= -i\frac {2r_+}{d-3}\partial_t\ , \nonumber \\
&& H_{-1}= \frac i{(d-3)r^{\frac d2-2}} e^{\frac {-(d-3)t}{2r_+}} \left(\Delta^{1/2} \partial_r
+\frac {r_+ (2r^{d-3}-r_+^{d-3})}{\Delta^{1/2}} \partial_t \right)\ , \nonumber \\
&& {\cal H}^2= \frac 1{(d-3)^2 r^{d-4}} \left[ \partial_r (\Delta \partial_r\ ) -\frac {r_+^{2d-4}}{\Delta} \partial_t^2 \right]\ .
\end{eqnarray}

\end{document}